\documentstyle[12pt]{article}
\font\grande=cmr10 scaled \magstep4
\font\medio=cmr10 scaled \magstep2

\def\laq{\raise 0.4 ex \hbox{$<$}\kern -0.8 em\lower 0.62 ex\hbox{$\sim$}}
\def\gaq{\raise 0.4 ex \hbox{$>$}\kern -0.7 em\lower 0.62 ex\hbox{$\sim$}}

\def\vev#1{\langle {#1}\rangle}
\def\laq{\raise 0.4ex\hbox{$<$}\kern -0.8em\lower 0.62
ex\hbox{$\sim$}}
\def\gaq{\raise 0.4ex\hbox{$>$}\kern -0.7em\lower 0.62
ex\hbox{$\sim$}}

\def\AJ{{\it Astrophys. J.} }
\def\AJL{{\it Ap. J. Lett.} }

\def\CQG{{\it Class. Quantum Gravity} }

\def\MPL{{\it Mod. Phys. Lett.} }

\def\PM{{\it Philos. Mag. } }
\def\PL{{\it Phys. Lett.} }
\def\PR{{\it Phys. Rev.} }
\def\PRL{{\it Phys. Rev. Lett.} }

\def\RPP{{\it Rep. Progr. Phys.}}

\def\UFN{{\it Usp. Fiz. Nauk} }

\def\SPU{{\it Soviet. Phys. Usp.} }

\begin{document}
\bibliographystyle {unsrt}
\newcommand{\pa}{\partial}

\titlepage
\begin{flushright}
DF/IST--3.98 \\
March 1999
\end{flushright}
\vspace{15mm}

\begin{center}
{\bf Primordial Magnetic Fields via Spontaneous Breaking of Lorentz 
Invariance}\\
\vspace{5mm}
{\grande }
\vspace{10mm}
{\bf O. Bertolami and D. F. Mota}\\
{\em Departamento de F\'{\i}sica,
Instituto Superior T\'ecnico}\\ 

{\em Av. Rovisco Pais, 1096 Lisboa Codex, Portugal } \\
\end{center}

\vspace{10mm}
\centerline{\medio  Abstract}
\vspace{1cm}

\noindent
Spontaneous breaking of Lorentz invariance compatible 
with observational limits may realistically take place in 
the context of string theories, possibly endowing the photon with a mass. 
In this process the conformal symmetry of the
electromagnetic action is broken allowing for 
the possibility of generating large scale ($\sim Mpc$) magnetic 
fields within inflationary scenarios. We show that for reheating
temperatures safe from the point of view of the gravitino and moduli problem,
$T_{RH}~\laq~10^{9}~GeV$ for $m_{3/2} \approx 1~TeV$, the strength of the 
generated seed fields is, in our mechanism, consistent with amplification by  
the galactic dynamo processes and can be even  
as large as to explain the observed galactic magnetic 
fields through the collapse of protogalactic clouds.\\

\vspace{1cm}

\noindent
PACS number(s): 98.80 Cq, 98.62 En

\vfill

\newpage

\setcounter{equation}{0}
\setcounter{page}{2}

\setlength{\baselineskip}{0.8cm}

\begin{flushleft}
{\bf 1. Introduction}
\end{flushleft}

Coherent magnetic fields are observed over a wide range of scales in the 
Universe from the Earth, Solar System and stars to galaxies and 
clusters of galaxies \cite{kronberg}.

Magnetic fields play an important role in a variety of astrophysical 
processes. For instance, the galactic field affects the dynamics of the 
galaxy as it confines cosmic rays, influences the dynamics 
of compact stars and 
the process of star formation \cite{zeldovich}. Large scale magnetic 
fields are also quite important in quasars, active galactic nuclei and in 
intercluster gas or rich clusters of galaxies.  

The current estimates for the magnetic field of the Milky Way and nearby 
galaxies is $B \sim 10^{-6}$ G, which are supposed to be coherent 
over length scales comparable to the size of the galaxies themselves 
\cite{kronberg}. The origin of the astrophysical mechanisms responsible for 
these galactic fields is however poorly understood. 

Possibly the most plausible explanation for the observed galactic magnetic 
field involves some sort of dynamo effect \cite{zeldovich, parker}. In 
this mechanism, turbulence generated by the differential rotation of the 
galaxy enhances exponentially, via non-linear processes, a seed magnetic 
field up to some saturation value that corresponds to equipartition between 
kinetic and magnetic energy. A galactic dynamo mechanism along these lines 
can enhance a seed magnetic field by a factor of several orders of magnitude. 
Indeed, if one assumes that the 
galactic dynamo has operated during about 10 Gyr, then a seed magnetic field 
could be amplified by a factor of $e^{30}$ corresponding to about 30 
dynamical timescales or complete revolutions since the galaxy has formed. 
Hence the observed galactic magnetic fields at present may, via the dynamo 
amplification, have its origin in a seed magnetic field of 
about $B \sim 10^{-19}G$.

Naturally, the galactic magnetic fields can emerge directly from the 
compression of a primordial magnetic field, in the collapse of the
protogalactic cloud. In this case, it is required a seed magnetic field of 
$B \sim 10^{-9}G$ over a comoving scale $\lambda \sim Mpc$, the comoving size 
of a region which condenses to form a galaxy. Since the Universe through most 
of its history has behaved as a good conductor it 
implies that the evolution of any primeval cosmic magnetic field will 
conserve magnetic flux \cite{turner, ahonen}. Therefore, the ratio denoted 
by $r$, of the energy 
density of a magnetic field $\rho_B = {B^2 \over 8\pi}$ relative to the 
energy density of the cosmic microwave background radiation 
$\rho_\gamma = {\pi^2 \over 15} T^4$ remains approximately constant and 
provides a invariant measure of 
magnetic field strength.

At present, for galaxies $r\sim 1$, so pregalactic magnetic fields of about 
$r \approx 10^{-34}$ are required if one invokes dynamo 
amplification processes, and 
$r\approx 10^{-8}$ if the observed galactic magnetic fields are created from 
the compression of the primordial magnetic field in the collapse of the 
protogalactic cloud without the help of dynamo type processes.

A number of proposals have been put forward to explain the way a 
primordial magnetic field could be generated 
(see \cite{enqvist} for a review). 
In many of these proposals, changes in the nature of 
the electromagnetic 
interaction at the period of inflation \cite{turner, garretson} are 
involved together with the process of structure formation \cite{turner,davis}. 
Other proposals invoke collision of phase transition bubbles \cite{kibble}, 
fluctuating Higgs field gradients \cite{vashaspati}, 
superconducting cosmic strings \cite{dimopoulos} and non-minimal coupling 
between electromagnetism and gravity via the Schuster-Blacket relation 
\cite{opher}. 
In this work we shall describe a mechanism to produce a seed magnetic field 
that is based on a possible violation of Lorentz invariance in solutions of 
string field theory and that uses inflation for amplification of 
quantum fluctuations of the electromagnetic field. We show that these 
fluctuations are
compatible with both galactic dynamo mechanisms and  
protogalactic cloud collapse scenario.

Invoking a period of inflation to explain the creation of 
seed magnetic fields is a quite attractive suggestion as inflation 
provides the means of generating large-scale phenomena from 
microphysics that operates on subhorizon scales. More concretely, inflation, 
through de Sitter-space-produced quantum fluctuations, provides the means of 
exciting the electromagnetic field allowing for an increase of the magnetic 
flux before the Universe gets filled with a highly conducting 
plasma. Furthermore, via a mechanism akin to the superadiabatic 
amplification, long-wavelength modes, for which $\lambda \gaq\ H^{-1}$, are 
during inflation and reheating enhanced. It is certainly quite 
interesting that inflation can play, for generating primordial magnetic
fields, the same crucial role it plays in solving the problems of
initial conditions of the cosmological standard model.

However, as pointed out by several authors \cite{turner, garretson}, 
it is not possible to produce the required seed magnetic fields 
from a conformally invariant theory as it 
happens with the usual U(1) gauge theory. The reason being that, in a 
conformally invariant theory, the magnetic field decreases as 
$a(t)^{-2}$, where $a(t)$ is the scale factor of the Robertson-Walker 
metric, and during inflation, the total energy density in the Universe is 
constant, so the magnetic field energy density is strongly suppressed, 
yielding $r = 10^{-104}\lambda_{Mpc}^{-4}$, which is far too low for a seed 
field candidate. It then follows that conformal invariance of 
electromagnetism must be broken.

In the context of string theories, conformal invariance may be broken 
actually due to the possibility of spontaneous breaking of the Lorentz 
invariance \cite{kostelecky} (this breaking can also lead to
the breaking of CPT symmetry \cite{potting}). 
This possibility arises explicitly from solutions 
of string field theory, at least 
for the open type I bosonic string, as interactions are cubic 
in the string field and these give origin in the static field theory potential
to cubic interaction terms of the type $SSS$, $STT$ and $TTT$, where 
$S$ and $T$ denote scalar and tensor fields. The way Lorentz invariance 
may be broken can be seen, for instance, from the  
static potential involving the tachyon and a generic vector field as can be 
explicitly computed \cite{kostelecky}:

\begin{equation}
\label{potential}
V(\varphi,A_{\mu}, ...) =  - {\varphi^2 \over 2 \alpha'} + a g \varphi^3 
+ b g  \varphi V_{\mu} V^{\mu} + ... ,
\end{equation}
$a$ and $b$ being order one constants and $g$ the on-shell three-tachyon 
coupling.

The vacuum of this model is clearly unstable and this instability 
gives rise to a mass-square term for the vector field that is proportional 
to $\vev{\varphi}$. If $\vev{\varphi}$ is negative, 
then the Lorentz symmetry is spontaneously broken as the vector field
can acquire itself a non-vanishing vacuum expectation values\footnote{
Actually, it is known that negative quadratic mass states corresponding to 
tachyonic solutions are admitted in the representation space of massive vector 
particles implying in a violation of 
the rotational sector of Lorentz invariance \cite{ahluwalia}.}. 
This mechanism can give rise to vacuum expectation values to tensor 
fields inducing for the fields that do not acquire vacuum expectation values,
such as the photon, to mass-square terms proportional
to $\vev{T}$ (this possibility has been briefly discussed in \cite{kps}). 
Hence, one should expect from this mechanism, terms for the 
photon such as $\vev{T}A_{\mu}A^{\mu}$, 
$\vev{T_{\mu \nu}}A^{\mu}A^{\nu}$ and so on. Naturally, these terms break
explicitly the conformal invariance of the electromagnetic action.

Observational constraints on the breaking of the Lorentz invariance 
arise from measurements of the quadrupole splitting time 
dependence of nuclear Zeeman levels along Earth's orbit, the so-called 
Hughes-Drever experiment \cite{hrbl, drever}, and have been performed 
over the  years \cite{prestage, lamoreaux}, the most recent 
one indicating that $\delta < 3 \times 10^{-21}$ \cite{chupp}.
Bounds on the violation of momentum conservation and  
existence of a preferred reference frame can be also 
extracted from limits on the
parametrized post-Newtonian parameter $\alpha_{3}$ obtained
from the pulse period of pulsars \cite{will} and millisecond  
pulsars \cite{bell}.
This parameter vanishes identically in general relativity and the  
most recent limit $|\alpha_{3}| < 2.2 \times 10^{-20}$ obtained from  
binary pulsar systems \cite{bd} implies the Lorentz symmetry is unbroken  
up to this level. These limits indicate that if the Lorentz invariance is
broken then its violation is suppressed by powers of energy over the string
scale. Similar conclusions can be drawn for putative violations of the
CPT symmetry \cite{potting}.

In order to relate the theoretical possibility of spontaneous breaking of
Lorentz invariance to the observational limits discussed above 
we parametrize
the vacuum expectation values of the Lorentz tensors in the following way:

\begin{equation}
\label{parametrization}
<T> = m_{L}^2 \left({E \over M_{S}}\right)^{2l}~,
\end{equation}
where $m_L$ is a light mass scale when compared to string typical
energy scale, $M_{S}$, presumably $M_{S} \approx 
M_{P}$; $E$ is the temperature of the Universe in a given period and $2l$ is 
a positive integer. Given that the expansion of 
the Universe is adiabatic, we shall further replace in what follows 
the temperature of
the Universe by the inverse of the scale factor (the proportionality constant
is absorbed in the yet unspecified light mass scale, $m_{L}$). 
Notice that parametrization 
(\ref{parametrization}) used here is somewhat different than the 
ones used in previous work \cite{potting,bertolami1,bertolami2}. 


\begin{flushleft}
{\bf 2. Generation of Seed Magnetic Fields}
\end{flushleft}

We are going to consider spatially flat Friedmann-Robertson-Walker 
cosmologies, where the stress tensor is described by a perfect fluid 
with an equation of 
state $p = \gamma \rho$. The metric in the conformal time, $\eta$, is given by:

\begin{equation}
\label{metric}
g_{\mu \nu} = a(\eta)^2 diag(-1,1,1,1)~~,
\end{equation}
where $a(\eta)$ is the scale factor.

The present value of the Hubble parameter is written as 
$H_0 = 100 \ h_0~km~s^{-1}~Mpc^{-1}$ and the present Hubble radius is 
$R_0 = 10^{26}\ h_0 ^{-1} \ m$, where $0.4\leq h_0 \leq 1$. 
We shall assume the Universe has gone through a period of 
exponential inflation at a scale $M_{GUT}$ and whose
associated energy density is given by $\rho_I \equiv M_{GUT}^{4}$. The details 
of this de Sitter phase are not relevant and will play, as discussed in 
\cite{turner}, no role in our mechanism for generating a primordial
seed magnetic field. From the Friedmann equation, 
$H_{dS} = ({8 \pi G \over 3}\rho_I)^{1/2} =
({8\pi \over 3})^{1/2}~{M_{GUT}^2 \over M_P}$, where $M_P$ is 
the Planck mass.

From our discussion on the breaking of Lorentz 
invariance we consider for simplicity only a single term, namely
$\vev{T}A_{\mu}A^{\mu}$, from which implies the following 
Lagrangian density for the photon:

\begin{equation}
\label{lagrangian}
{\cal L} = - \frac{1}{4} F_{\mu \nu} F^{\mu \nu} + M_L^2 a^{-2l} A_\mu A^\mu ,
\end{equation}
where $M_L^2 \equiv {m_L^2 \over M_p^{2l}}$.

The field strength tensor, $F_{\mu \nu}$, is given by

$$
\displaylines{F_{\mu \nu} = a(\eta)^2 \pmatrix{0&-E_x&-E_y&-E_z\cr E_x&0&B_z
&-B_y\cr E_y&-B_z&0&B_x\cr E_z&B_y&-B_x&0}\cr} 
$$
and it satisfies the Bianchi identity

\begin{equation}
\label{bianchi}
\partial_{\mu}F_{\lambda \kappa} + \partial_{\lambda}F_{\kappa \mu}+ 
\partial_{\kappa}F_{\mu \lambda} = 0 
\end{equation}
as well as the equation of motion for the photon field
 
\begin{equation}
\label{eqmotion}
\nabla^{\mu} F_{\mu \nu} + M_L ^2 a^{-2l}A_{\nu} = 0~~.
\end{equation}

Eqs. (\ref{bianchi}) and (\ref{eqmotion}) can be then explicitly 
written as:

\begin{equation}
\label{eqmotionex}
{1 \over a^2} {\partial \over \partial \eta} a^2 \vec B + \vec \nabla \times 
\vec E = 0~,
\end{equation}

\begin{equation}
\label{propertyex}
{1 \over a^2} {\partial \over \partial \eta} a^2 \vec E - \vec \nabla \times 
\vec B - {n \over \eta^2} {\vec A \over a^2} = 0~, 
\end{equation}

\noindent
where $n \equiv - \eta ^2 M_L^2 a ^{-2l+2}$ and $\vec A$ is the vector 
potential.

Taking the curl of Eq. (\ref{propertyex}) and using Eq. (\ref{eqmotionex}) we 
obtain the wave equation for the magnetic field:

\begin{equation}
\label{motionB}
{1 \over a^2} {\partial^{2} \over \partial \eta^2} a^2 \vec B - \nabla ^2 
\vec B + {n \over \eta^2} \vec B = 0~.
\end{equation}

The corresponding equation for the Fourier components of 
$ \vec B$ is given by:

\begin{equation}
\label{motionF}
\ddot {\vec F_k} + k^2 \vec F_k + \frac{n}{\eta ^2} \vec F_k = 0~,
\end{equation}
where the dots denote derivatives according to the conformal time and 

\begin{equation}
\label{Ffourier}
\vec F_k(\eta) \equiv a^2 \int{ d^3x e^{i \vec k . \vec x} \vec 
B( \vec x, \eta)}~,
\end{equation}
$\vec F_k$ being a measure of the magnetic flux associated with the
comoving scale $ \lambda \sim k^{-1} $. It follows that the energy 
density of the magnetic field is given by 
$\rho_B(k) \propto {|\vec F_k|^2 \over a^4}$.

For modes well outside the horizon, $a \lambda >> H^{-1}$ or $|k \eta| << 1$, 
solutions of Eq. (\ref{motionF}) are given in terms of the conformal time:

\begin{equation}
\label{Feta}
| \vec F_k| \propto \eta ^{m_{ \pm}}
\end{equation}
where
\begin{equation}
\label{exponent}
m_{ \pm} = \frac{1}{2} \left[1 \pm \sqrt{1-4n}\right].
\end{equation}

In order to estimate $n$ we consider the different phases of evolution
of the Universe. By requiring that $n$ is not a 
growing function of conformal time, it follows that $n$ has to be 
either a constant 
or that $2l$ is negative, which is excluded by our assumption 
(\ref{parametrization}). Hence for different phases of evolution 
of the Universe:

\noindent
(I) Inflationary de Sitter (dS) phase, where $a( \eta) \propto - 
{1 \over \eta H_{dS}}$, it follows that $l = 0$ and

\begin{equation}
\label{sittern}
n = - \frac{M_{dS}^2}{H_{dS}^2}~~,
\end{equation}
where we denote the light mass $M_L$ by the index of the relevant phase 
of evolution of the Universe.

\noindent
(II) Phase of Reheating (RH) and Matter Domination (MD), where $a( \eta) 
\propto \frac {1}{4} H_0^2 R_0^3 \eta ^2$, yields from the 
condition $n$ is a constant that $2l = 3$ and

\begin{equation}
\label{MDRHn}
n = - \frac{4M_{MD}^2}{H_0^2R_0^3}~~. 
\end{equation}

\noindent
(III) Phase of Radiation Domination (RD), where $a( \eta) \propto 
H_0 R_0^2 \eta $, from which follows that $l = 2$ and

\begin{equation}
\label{RDn}
n = - \frac{M_{RD}^2}{H_0^2R_0^4}~~. 
\end{equation}
It is clear that in this case last case $n \ll 1$.

We mention that, we could have obtained the same results comparing Eq. 
(\ref{motionF}) to the corresponding equation of the Fourier modes of a 
scalar field, coupled non-minimally to gravity through the term 
$\frac{1}{2} \xi R \phi ^2$, where 
$R = 6 {\ddot a \over a^3}$ is the Ricci scalar. 
The relevant Lagrangian density is:

\begin{equation}
\label{scalarf}
{\cal L}_{\phi} = - {1 \over 2} 
\partial_\mu \phi \partial^\mu \phi - {1 \over 2} \xi R \phi ^2~~.
\end{equation}

In terms of the $k$-th Fourier component of the combination 
$w = a \phi$, the equation of motion of a non-minimally 
coupled scalar field can be written as \cite{turner}: 

\begin{equation}
\label{Wfourier}
\ddot w_k + k^2w_k + \frac{n_\phi}{\eta ^2}w_k =0~~, 
\end{equation}
where $n_\phi = \eta ^2 (6 \xi -1) {\ddot a \over a}$ and as before 
$\rho_{\phi}(k) \propto {|w_k|^2 \over a^4}~$.

Notice that the correspondence between $w_k$ and the 
components of $\vec F_k$ implies, for instance, that if 
$\xi = \frac{1}{6}$ then the components of 
$A_\mu$ behave like a scalar field conformally coupled and if $\xi = 0$ then 
the components of $A_\mu$ behave like a scalar field minimally coupled.
Furthermore, a  correspondence between $w_k$ and $\vec F_k$, 
requires that $n = n_{\phi}$ implying in the following 
condition:

\begin{equation}
\label{compare}
\ddot a + \frac {M_L^2}{(6 \xi -1)}a^{-2l+3} = 0~~.
\end{equation}

It is easy to show that for different phases of evolution of the Universe one 
can obtain, via Eq. (\ref{compare}), essentially the same results for $n$ 
as the ones we have obtained from requiring that $n$ is constant.

Hence for each phase of evolution of the Universe we find
from the corresponding values of $l$ and associated conditions to $M_L$ the 
following behaviour for $|\vec F_k|$:

\noindent
(I) de Sitter Phase 

\begin{equation}
\label{sitterF}
|\vec F_k| \propto a^{- m_{ \pm}}~,
\end{equation}
where

\begin{equation}
\label{sitterm}
m_{ \pm dS} = \frac{1}{2} \left[1 \pm \sqrt{1+
\left({2M_{dS} \over H_{dS}}\right)^2~}\right]
\end{equation}
and 

\begin{equation}
\label{densityds}
\rho_{BdS} \propto a^{-2m_{ \pm dS}-4}~~.
\end{equation}

\noindent
Notice that the most relevant exponent is given by $m_{-dS}$, as it 
corresponds to the fastest growing solution for $ \vec F_k$.

\noindent
(II) Phase of Reheating and Matter Domination

\begin{equation}
\label{rhmdF}
|\vec F_k| \propto a^{ \frac{1}{2} m_{ \pm}},
\end{equation}
where

\begin{equation}
\label{rhmdm}
m_{ \pm RH} = {1 \over 2} \Biggl[1 \pm \sqrt{1 +
16~{M_{RH}^2 \over H_{0}^2R_0^3}~~}\Biggr]~~,
\end{equation}
and therefore

\begin{equation}
\label{densityrhMD}
\rho_B \propto a^{m_{ \pm MD}-4}~~.
\end{equation}

\noindent
The relevant exponent here is $m_{+RD}$, as it corresponds to the fastest 
growing solution for $\vec F_k$.

\noindent
(III) Phase of Radiation Domination

\begin{equation}
\label{rdF}
|\vec F_k| \propto a^{ m_{ \pm}}~~,
\end{equation}
where

\begin{equation}
\label{rdm}
m_{ \pm RD} = 1 , 0
\end{equation}
from which follows that

\begin{equation}
\label{densityRD}
\rho_B \propto a^{-2} \ or \ \rho_B \propto a^{-4}~~.
\end{equation}
As expected, for $m_{-RD}$ one obtains that $\rho_B \propto a^{-4}$.

These results enable us to estimate strength of the primordial 
magnetic field. Assuming the Universe has gone through a period
of inflation at scale $M_{GUT}$ and that fluctuations of the electromagnetic 
field have come out from the horizon when the Universe had gone through about
$55$ $e$-foldings of inflation, then in terms of $r$ \cite{turner}:

\begin{equation}
\label{ratio}
r \approx (7 \times 10^{25})^{-2(p+2)} \times (\frac{M_{GUT}}{M_{P}})
^{4(q-p)/3} \times (\frac{T_{RH}}{M_{P}})^{2(2q-p)/3} 
\times (\frac{T_{*}}{M_{P}})^{- 8q/3} \times 
\lambda_{Mpc}^{-2(p+2)},
\end{equation}
where $T_{*}$ is the temperature at which plasma effects become dominant,
that is, the temperature when the Universe first becomes a good conductor
and that can be estimated from details of the reheating process
$T_{*} = min \{(T_{RH}M_{GUT})^{ \frac{1}{2}};(T_{RH}^2 M_{p})^{ 
\frac{1}{3}}\}$ \cite{turner}; for 
$T \leq T_{*}$, $ \rho _B$ necessarily evolves as 
$ \propto a^{-4}$. For the reheating temperature we assume either a
poor or a quite efficient reheating, $T_{RH} = \{ 10^9~GeV ; M_{GUT} \}$, 
see however the discussion below. 
Finally, $p \equiv m_{-dS} = {1 \over 2} \left[1 - \sqrt{1+
\left({2M_{dS} \over H_{dS}}\right)^2}\right]$ and $q \equiv m_{+ RH} 
= {1 \over 2} \Biggl[1 + \sqrt{1+ 16 {M_L^2 \over H_0^2 R_0^3}}\Biggr]$ 
are the fastest growing solutions 
for $\vec F_k$ in the de Sitter and Reheating phases, respectively. 

In order to obtain numerical estimates for $r$ we have to compute $M_L$. 
At the de Sitter phase we have that 
$M_{dS} = m_{dS}$. As we have seen $m_L$ is a 
light energy scale when compared to $M_P$ and 
$M_{GUT}$, the energy scale of the de Sitter phase. Thus in order
to estimate $M_{L}$ at the de Sitter phase 
we introduce a parameter, $\chi$, such that $m_{dS} = \chi M_{GUT}$ 
and $\chi \ll 1$. 

At the matter domination phase, we have to impose that the mass term 
$M_{MD} = m_{MD} ({T_{\gamma} \over M_{p}})^{l}$, $T_{\gamma}$ 
being the temperature of the cosmic background radiation at about 
the recombination time, is consistent with the 
present-day limits of experiments and observations to the photon mass. Thus, 
at the matter domination phase, we have to satisfy the condition, 
$M_{MD} \leq m_{\gamma}$ which implies that 
$m_{MD} ({T_{\gamma} \over M_{P}})^{3/2} \leq 3 \times 10^{-36}~GeV$ 
\cite{chibisov}, following that $m_{MD} \leq 7.8 \times 10^{4}~GeV$.
A more stringent bound on $m_{MD}$ could be obtained from the limit
$m_{\gamma} \leq 1.7 \times 10^{-42}~h_{0}~GeV$ arising from the
absence of rotation in the polarization of light of distant galaxies 
due to Faraday effect \cite{carrol}.

In tables I, II and III we present our estimates for the ratio $r$
for different values of $M_{GUT}$. One can see that we obtain values that 
are in the range $10^{-37} < r < 10^{-5}$. For 
$M_{GUT} = 10^{15},10^{16}~GeV$
one sees that a poor reheating and the lower values for $\chi$ 
tend to render $r$ 
too low even for an amplification via dynamo processes. On the other hand,
for $M_{GUT} = 10^{17}~GeV$ a quite efficient reheating leads for 
$\chi > 5 \times 10^{-2}$ to rather large primordial magnetic fields.
Actually, primordial fields greater than 
$3.4 \times 10^{-9}~(\Omega_{0} h_{50}^0)^{1/2}~G$\footnote{This constraint
is much more stringent than the ones arising from nucleosynthesis that imply,
at the end of nucleosynthesis when $T_{\gamma} \approx 0.01~MeV$, that
$B_{P} < 2 \times 10^{9}~G$ and that $\rho_{B} \le 0.28 \rho_{\nu}$
\cite{cheng}.}, where $\Omega_{0}$
is the density parameter at present and $h_{50}$ is the present value
of the Hubble parameter in units of $50~km~s^{-1}~Mpc^{-1}$, 
are ruled out as they lead to more anisotropies than the ones observed
in the microwave background radiation \cite{barrow}. Hence, our results
indicate that models with $M_{GUT} = 10^{17}~GeV$,
$\chi > 5 \times 10^{-2}$ and $T_{RH} = M_{GUT}$ should be disregarded.

However, an important issue is that in supersymmetric theories the 
reheting temperature is severely constrained in order to avoid that
gravitinos and moduli are not copiously regenerated in the post-inflationary
epoch. This is indeed a difficulty as as once regenerated beyond a 
certain density these particles dominate the energy density 
of the Universe or, if decay, have undesirable effects on nucleosynthesis and
lead to distortions of the microwave background. The relevant bounds are
the following (see \cite{bento} and references therein):

\begin{equation}
\label{bcaa}
T_{RH}~\le ~(2 - 6) \times
10^9~\hbox{GeV}~~~~\hbox{for}~~~~m_{3/2} = (1 - 10)~\hbox{TeV}~.
\end{equation}
Therefore it follows that for $M_{GUT} = 10^{15}~(10^{16})~GeV$, only for
$\chi > 5 \times 10^{-4}~(\chi > 5 \times 10^{-3})$ the generated 
seed magnetic 
fields are large enough. For $M_{GUT} = 10^{17}~GeV$, the generated seed 
fields, although rather small, can be still compatible with observations, 
for the chosen $\chi$ values, via the dynamo amplification process.

Finally, we mention that from the comparison with a non-minimally 
coupled scalar field, our results allow inferring, for example that for
$M_{GUT} = 10^{17}~GeV$, at 
the de Sitter phase $\xi = -0.02$ and $ \xi = -0.04$ for
$\chi = 5 \times 10^{-2}$ and $\chi = 6 \times 10^{-2}$, respectively. This
allows us to conclude that the growth in magnetic flux is analogous to 
the phenomenon of superadiabatic amplification, as the components of 
$A_{ \mu}$ behave like the scalar field with a negative mass-square term.
For the matter domination phase we find that $\xi = \frac{1}{6}$, actually 
the expected result.


\begin{flushleft}
{\bf 3. Conclusions}
\end{flushleft}

Galactic dynamo or protogalactic cloud collapse can 
explain the observed magnetic fields of galaxies at present
provided a primordial seed magnetic field is generated prior the Universe 
turns a good conductor and the magnetic flux turns into a conserved
quantity.

A particularly interesting proposal is to invoke inflation for explaining the 
origin of the primordial magnetic field. This is a quite good idea since 
inflation naturally relates microphysics with macrophysics as it
allows for amplification of quantum fluctuations of fields, and in particular
of the electromagnetic field, stretching fluctuations 
beyond the horizon prior
the Universe becomes a good conductor and locks the growth of electromagnetic
flux. However, this scenario requires the breaking of the conformal
invariance as otherwise the $r$ will be diluted to quite small
values $r < 10^{-104}\lambda_{Mpc}^{-4}$ at $Mpc$ scales. In Ref. 
\cite{turner} it was suggested, in
order to break the conformal symmetry of electromagnetism, 
to introduce in a somewhat {\it ad hoc} way terms
such as $RA_{\mu}A^{\mu}$, $R_{\mu \nu}A^{\mu}A^{\nu}$, etc, 
into the electromagnetic action. In this
work we have pointed out that the existence of string field theory solutions,
in the context of which Lorentz invariance
can be spontaneously broken, leads for the photon to terms 
like $\vev{T}A_{\mu}A^{\mu}$ $\vev{T_{\mu \nu}}A^{\mu}A^{\nu}$, 
hence breaking conformal symmetry. We then showed that this allows 
inflation to generate rather large seed magnetic fields.
   
Furthermore, we have demonstrated that the strength of the magnetic field 
produced by our mechanism is sensitive to 
the values of a light mass, $m_L$, (cf. Eq. (\ref{parametrization})), 
$M_{GUT}$ and the reheating temperature $T_{RH}$. 
Our results indicate that for rather diverse values of these 
parameters we can obtain values for $r$ that are consistent with amplification
via galactic dynamo or collapse of protogalactic clouds. 
For $M_{GUT} = 10^{15},10^{16}~GeV$ we find that 
poor reheating and $\chi = 5 \times 10^{-4}$ and $\chi = 5 \times 10^{-3}$ 
tend to give rise to too low $r$ 
values. For $M_{GUT} = 10^{17}~GeV$, very efficient reheating leads for
$\chi = 6 \times 10^{-2}$ 
to rather large primordial magnetic fields, these being
actually incompatible with upper limits derived from the study of the 
microwave background anisotropies. Since in supersymmetric theories the
reheating temperature is strongly constrained not to be greater than about
$10^{9}~GeV$ for ${\cal O}(TeV)$ gravitino and moduli masses, 
our results indicate that
for $M_{GUT} = 10^{15}, 10^{16}~GeV$ only by choosing 
$\chi > 5 \times 10^{-4}$ and $\chi > 5 \times 10^{-3}$, respectively, 
the generated seed fields are large enough for accounting the observations. 
There is actually no such a limitation for $M_{GUT} = 10^{17}~GeV$.


\clearpage

\begin{center}
{\bf Tables }
\end{center}

\begin{table}[!h]
\centering
\begin{tabular}{|l|l|l|l|l|l|} \hline\hline
${ \em \chi }$ & ${ \em p}$ & ${ \em q}$ & ${ \em T_{RH}(GeV)}$ 
& ${ \em T_{*}}(GeV)$ & ${ \em \log~r}$  \\ \hline\hline
$5 \times 10^{-4}$ & $-1.67$ & $1$ & $10^{9}$ & $10^{12}$ & $-37$  \\ \hline
$5 \times 10^{-4}$ & $-1.67$ & $1$ & $10^{15}$ & $10^{15}$ & $-31$  \\ \hline
$6 \times 10^{-4}$ & $-2.08$ & $1$ & $10^{9}$ & $10^{12}$ & $-21$  \\ \hline
$6 \times 10^{-4}$ & $-1.08$ & $1$ & $10^{15}$ & $10^{15}$ & $-13$  \\ \hline
\end{tabular}
\caption{Values of $r = \frac{\rho_B}{\rho_\gamma}$ at
$1~Mpc$ for $M_{GUT} = 10^{15}~GeV$}
\end{table}

\begin{table}[!h]
\centering
\begin{tabular}{|l|l|l|l|l|l|} \hline\hline
${ \em \chi }$ & ${ \em p}$ & ${ \em q}$ & ${ \em T_{RH}(GeV)}$ 
& ${ \em T_{*}}(GeV)$ & ${ \em \log~r}$  \\ \hline\hline
$5 \times 10^{-3}$ & $-1.67$ & $1$ & $10^{9}$ & $2.3 \times 
10^{12}$ & $-35$  \\ \hline
$5 \times 10^{-3}$ & $-1.67$ & $1$ & $10^{16}$ & $10^{16}$ & $-27$  \\ \hline
$6 \times 10^{-3}$ & $-2.08$ & $1$ & $10^{9}$ & $2.3 \times 10^{12}$ & 
$-18$  \\ \hline
$6 \times 10^{-3}$ & $-1.08$ & $1$ & $10^{16}$ & $10^{16}$ & $-9$  \\ \hline
\end{tabular}
\caption{Values of $r = \frac{\rho_B}{\rho_\gamma}$ at 
$1~Mpc$ for $M_{GUT} = 10^{16}~GeV$}
\end{table}

\begin{table}[!h]
\centering
\begin{tabular}{|l|l|l|l|l|l|} \hline\hline
${ \em \chi }$ & ${ \em p}$ & ${ \em q}$ & ${ \em T_{RH}(GeV)}$ & 
${ \em T_{*}}(GeV)$ & ${ \em \log~r}$  \\ \hline\hline
$5 \times 10^{-2}$ & $-1.67$ & $1$ & $10^{9}$  & $10^{13}$ & $-33$ \\ \hline
$5 \times 10^{-2}$ & $-1.67$ & $1$ & $10^{17}$ & $10^{17}$ & $-24$\\ \hline
$6 \times 10^{-2}$ & $-2.08$ & $1$ & $10^{9}$ &  $10^{13}$ & $-16$ \\ \hline
$6 \times 10^{-2}$ & $-1.08$ & $1$ & $10^{17}$  & $10^{17}$ & $-5$\\ \hline
\end{tabular}
\caption{Values of $r = \frac{\rho_B}{\rho_ \gamma}$ at
$1~Mpc$ for $M_{GUT} = 10^{17}~GeV$}
\end{table}

\clearpage


\end{document}